\documentclass[sigconf]{acmart}
\usepackage{multirow}
\usepackage{threeparttable}
\usepackage{subfigure}
\usepackage{array}
\usepackage[misc]{ifsym}
\usepackage[ruled,linesnumbered]{algorithm2e}  
\usepackage{url}
\usepackage{amsmath}  
\makeatletter
\renewcommand{\@thesubfigure}{\hskip\subfiglabelskip}
\makeatother
\AtBeginDocument{%
  \providecommand\BibTeX{{%
    \normalfont B\kern-0.5em{\scshape i\kern-0.25em b}\kern-0.8em\TeX}}}

\setcopyright{acmcopyright}

\begin{document}
\pagestyle{plain} 
\pagestyle{empty}
\copyrightyear{2019} 
\acmYear{2019} 
\acmConference[ACSAC '19]{2019 Annual Computer Security Applications Conference}{December 9--13, 2019}{San Juan, PR, USA}
\acmBooktitle{2019 Annual Computer Security Applications Conference (ACSAC '19), December 9--13, 2019, San Juan, PR, USA}
\acmPrice{15.00}
\acmDOI{10.1145/3359789.3359801}
\acmISBN{978-1-4503-7628-0/19/12}
\title{How to Prove Your Model Belongs to You:\\
A Blind-Watermark based Framework to Protect Intellectual Property of DNN}

\author{Zheng Li}
\orcid{0000-0002-4466-7523}

\affiliation{%
  \institution{School of Computer Science and Technology}
  \institution{Shandong University\country{China}}
}
\email{zheng.li@mail.sdu.edu.cn}

\author{Chengyu Hu}
\authornotemark[1]
\affiliation{%
   \institution{Key Laboratory of Cryptologic Technology and Information Security, Ministry of Education}
  \institution{School of Cyber Science and Technology}
  \institution{Shandong University\country{China}}
}
\email{hcy@sdu.edu.cn}

\author{Yang Zhang}
\affiliation{%
  \institution{CISPA Helmholtz Center for Information Security}
  \city{Saarland Informatics Campus\country{Germany}}
  %\country{France}
}
\email{yang.zhang@cispa.saarland}

\author{Shanqing Guo}
\authornote{Corresponding author.}
\affiliation{%
 \institution{Key Laboratory of Cryptologic Technology and Information Security, Ministry of Education}
 \institution{School of Cyber Science and Technology}
 \institution{Shandong University\country{China}}
}
\email{guoshanqing@sdu.edu.cn}

\renewcommand{\shortauthors}{Zheng Li et al.}

%%DS




\begin{abstract}
Deep learning techniques have made tremendous progress in a variety of challenging tasks, such as image recognition and machine translation, during the past decade.
Training deep neural networks is computationally expensive and requires both human and intellectual resources. Therefore, it is necessary to protect the intellectual property of the model and externally verify the ownership of the model. However, previous studies either fail to defend against the evasion attack or have not explicitly dealt with fraudulent claims of ownership by adversaries. 
Furthermore, they can not establish a clear association between the model and the creator's identity.

To fill these gaps, in this paper, we propose a novel intellectual property protection (IPP) framework based on blind-watermark for watermarking deep neural networks that meet the requirements of security and feasibility. Our framework accepts ordinary samples and the exclusive logo as inputs, outputting newly generated samples as watermarks, which are almost indistinguishable from the origin, and infuses these watermarks into DNN models by assigning specific labels, leaving the backdoor as the basis for our copyright claim. We evaluated our IPP framework on two benchmark datasets and 15 popular deep learning models. The results show that our framework successfully verifies the ownership of all the models without a noticeable impact on their primary task. Most importantly, we are the first to successfully design and implement a blind-watermark based framework, which can achieve state-of-art performances on undetectability against evasion attack and unforgeability against fraudulent claims of ownership. Further, our framework shows remarkable robustness and establishes a clear association between the model and the author's identity.

\end{abstract}

\maketitle
\section{Introduction}
Intellectual Property (IP) refers to the protection of creations of the mind, which have both a moral and commercial value. 
% IP is protected in law by, for example, patents, copyright, and trademarks, which enable people to earn recognition or financial benefit from what they invent or create.
IP is protected under the law framework in the form of, for example, patents, copyright, and trademarks, which enable people to earn recognition or financial benefit from what they invent or create.
However, it is not always easy to protect intellectual property.
% Counterfeiters and infringers are increasingly sophisticated. They often exploit procedural loopholes, proactively seek to invalidate legitimate patents and trademarks and deploy advanced techniques such as reverse and evasion technologies.
Nowadays, counterfeiters and infringers often exploit procedural loopholes 
and utilize advanced techniques, such as reverse engineering, to invalidate legitimate patents and trademarks.

% Nowadays, deep neural networks have achieved great success in more and more fields, ranging from computer vision to natural language processing and intelligent medical treatment, etc. 
Deep learning techniques have witnessed tremendous development during the past decade, and are adopted in various fields ranging from computer vision \cite{CLXJYF17,HZRS16,HZRS162,HZCKWWAA17.0,SZ14.0,SLJSRAEVR15} to natural language processing \cite{XDHSSSYZ16.0,CWBKKKn,WSCLNMKCGMKSJKGKKKSKPWYSRRVCHD16.0}.
% However,  it  is  a  non-trivial  task  to  build  a  deep  learning model, especially a production-level model. We need to utilize human  expertise,  powerful  computing  resources,  and  large-scale datasets to train a high-performance model. For most of the common users like individuals and small enterprises, it is unmanageable  to  train  their  own  deep leanring  models. Therefore, the resultant models are considered as the IP of the creator, and requires the protection to maintain the creator's competitive advantage.
However, they are facing serious deep learning privacy and security  problems \cite{SZHFB18.0,HM18.0}.
It is a non-trivial task to build an effective model, especially at a production level.
Extensive computing power, large datasets, and human expertise are required.
Therefore, protecting the intellectual property of a model is essential to maintain the creator's competitive advantage.

% Discussions on the need for intellectual property protection for deep learning models have already begun around the world, and on November 1, 2018, the European Patent Office (EPO) review guidelines on the patentability of artificial intelligence and machine learning technologies have entered into force. 
The necessity of intellectual property protection for deep learning models has raised attention worldwide.
On November 1, 2018, the European Patent Office reviewed guidelines on the patentability of AI and machine learning technologies \cite{AI_patent,EPO_guidelines}. 
% The first step in intellectual property protection in the field of deep learning has been taken, and China and the United States are also formulating relevant policies. 
Other major political entities are also taking steps to formulate relevant policies.
% There is no doubt that the intellectual property protection of deep neural networks has begun. 
% However, it is almost impossible in practice: If you have a machine learning patent and you think somebody is infringing, there is almost no way to determine how their algorithm or machine learning works until you sue them, go for discovery, get access, get the information, and then figure it out --- which can be very expensive and risky. Followed the current rules of intellectual property protection, If you claim that someone abuses your model, you have the burden of proof, which means that you must provide evidence that he/she is more than 50\% likely to deploy your licensing model without permission.
However, protecting the IP of models faces difficulties.
A model owner can only rely on the legal system, the process of which is lengthy and expensive.
And following the current rules of IP protection, %If you claim that someone abuses your model, you have the burden of proof, which means that you must provide evidence that he/she is more than 50\% likely to deploy your licensing model without permission.
the proof of model ownership requires technical means.
% To this end, watermarking protection technology has been proposed, which has already been widely applied in protecting the intellectual property of multimedia content. 

To this end, digital watermark which has been widely applied to protect multimedia content is introduced to the IP protection of deep learning models. %Embedding digital watermarks into deep neural models is a key enabler for reliable technology transfer. % It aims to hide a watermark in the signal or copyright, including audio, video, image, etc. 
% From multimedia filed to DNN filed, the requirements for an effective watermarking methodology for DNN models are shown in Table I. 
An effective watermarking mechanism needs to satisfy multiple requirements including \textbf{Fidelity}, \textbf{Effectiveness}, \textbf{Integrity}, \textbf{Security}, \textbf{Legality} and \textbf{Feasibility}.
% However, practice has shown that none of the existing methods for watermark protection of DNN's intellectual property rights can meet all of the above requirements. The extension of watermarking protection technology to deep learning is still in its infancy.
However, none of the existing methods for watermarking deep learning models can meet all of the above requirements. 
In another way, the extension of watermarking protection technology to deep learning is still in its infancy.
%%%%%%%%%%%%%%%%
\subsection{Related Works}
Researchers have proposed approaches facilitating watermarking deep learning model for protecting intellectual property. 

Uchida et al. \cite{UNSS17} proposed a framework to watermark models for the first time in a white-box way. They assumed the owner can access the target model directly, including the model parameters, to verify the ownership. However, the stolen model is usually deployed as a remote service, which indicates that the model owner is actually unable to access the model parameters. %Merrer et al.\cite{merrer2017adversarial} proposed a watermark method using adversarial examples\cite{szegedy2013intriguing} for the first time in black-box way. An adversarial example is a sample sample created by adding a small perturbation to the sample so that the model mispredicts the resulting sample. Then they embeds them in the decision boundary of the original model by fine-tuning the DNN with the adversarial examples.

Rouhani et al. \cite{RCK18.0} proposed a watermark methodology that meets both the white-box and black-box requirements. They selected a pair of random images and random labels as the watermark, which is also called key samples. Zhang et al. \cite{ZGJWSHM18} proposed a similar watermarking method while they employed other multiple types of watermarks. Adi et al. \cite{ABCPK18} chosen a set of abstract images with pre-defined labels as a watermark. Guo et al. \cite{GP18} proposed a digital watermark technique by adding a  message marks associated with the signature to the original images as the watermark. %In the verification procedure, the model owner will issue prediction queries of watermarks and evaluate the agreement with the pre-defined labels of the watermarks. 
One obvious drawback is that the distribution of their key samples is distant from the origin. An attacker can easily build a detector to evade identification by detecting the key samples, thus avoiding the detection of model theft. Another attack is that the above watermarking methods are also vulnerable to the threat of fraudulent claims because the feature distribution of the key samples is significant and striking, an attacker can easily build a set of fake samples, coincidentally making the model behave as if it were real. We discuss the two vulnerabilities as our main motivations in section \ref{motivation}.

Namba et al. \cite{NSn} proposed a watermarking method that can defend against evasion attacks. They selected a set of original samples as a watermark from the training set with label change.  Although this approach is promising, it is as incapable of establishing a clear association between the model and the creator's identity as most of the existing methods \cite{ABCPK18,UNSS17,RCK18.0}.

\subsection{Our Contribution}
Therefore, we propose a novel IPP framework based on blind-watermark for watermarking deep neural networks that meets all the requirements of an effective watermarking mechanism. Our contributions in this paper are three-fold:
\begin{itemize}
  \item We propose the first blind-watermark based IPP framework aiming to generate the key samples of which the distribution is similar to the original samples, and clearly associate the model with an actual creator's identity.    
  \item We implement a prototype of our IPP framework and evaluate it on two benchmark image datasets and 15 popular classification DNN models.
  Extensive experiments show that our framework is effective to verify the ownership without significant side effects on primary tasks.
  \item We conduct extensive empirical validations to show that our IPP framework achieves state-of-art performances on undetectability, unforgeability, and robustness against several forms of attacks. 
  %\item [4)]
  %For the first time in the field of intellectual property protection for DNN, we emphasized and utilized the weaknesses of DNN --- the ininterpretability and the over-expressivity, and turn the weakness into the strength to leave backdoor in DNN model.
\end{itemize}

\section{Background}

In this section, we introduce the relevant background knowledge about deep neural networks and digital watermarks, which are closely related to our work.

\subsection{Deep Neural Networks (DNNs)} 

Deep neural networks are a crucial component of state-of-the-art artificial intelligence services, showing a level of exceeding humans in various tasks such as visual analysis, speech recognition, and natural language processing. They have dramatically changed the way we conceive software and quickly became a universal technology, and more importantly, it is significantly better than the most advanced machine learning algorithms previously used in these areas. 

Although deep neural networks have made significant progress in various fields, it is still a non-trivial task to build deep learning models, especially a production-level model. We need to utilize (1) a large-scale labeled training dataset that can completely cover potential scenarios. (2) a lot of computing power, exceptionally high-performance devices such as GPU, TPU, etc. (3) long-term training to update the parameters of the neural network. (4) the corresponding domain expertise and engineering knowledge to design the network structure and select the hyperparameters. Consequently, building a well-trained model constitutes an important part of the owner's IP, and it is essential to design an intellectual property protection technique to maintain the owner's competitive advantage and economic benefits. 

\subsection{Digital Watermark}
The digital watermark is a brand-new information hiding technology in the last twenty years. It is to embed the identification information (i.e., a digital watermark) directly into digital carriers (including multimedia, documents, software, etc.) without affecting the characteristics of the original. %The watermarking process consists of two steps: embedding and verification. In the first step, the embedding algorithm embeds the watermark in the digital carrier, which is the data to be protected. In the second step, we verify the ownership of the carrier by whether the watermark can be extracted from the carrier.

%A digital watermark can typically be divided into non-blind watermark and blind watermark, the former refers that the original cover signal is required during the detection process, and the latter indicates that the original carriers are not needed during the verification process to detect the watermark. 
Digital watermark can typically be divided into non-blind watermark and blind watermark, the former refers that it is perceived or noticed by a Human Visual System (HVS). In contrast, blind watermark is usually invisible or imperceptible. Due to the feature distribution of key samples generated by the existing watermark method varies greatly, previous studies based on non-blind watermark either fail to defend against the evasion attack or have not explicitly dealt with fraudulent claims of ownership by adversaries. Therefore, we need to design a new watermark framework based on blind-watermark to embed the watermark into the neural network model.

\begin{figure}[htbp]
\centerline{\includegraphics[width=7cm, height=3.5cm]{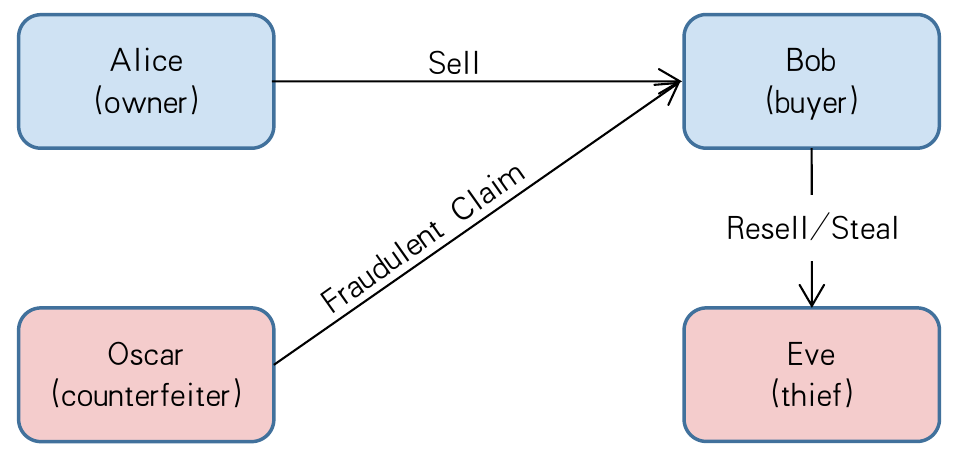}}
\caption{ The threat of security and legality}
\label{fig:motivation}
\end{figure}
\section{Motivation}\label{motivation}
In addition to the above limitations of previous works we addressed, in this section, we mainly discuss the two forms of attack: evasion attack and fraudulent claims of ownership. Figure \ref{fig:motivation} illustrates the threat.
\subsection{Security: Evasion Attack}
Suppose there is a model owner Alice, a thief Eve, and a model buyer Bob. Two possible scenarios can lead to evasion attack: (1) Eve has stolen the model in some way, (2) Bob has resold the model without Alice's authorization. Both of these behaviors are detrimental to the owner's interests. If the model is well-watermarked, the model owner Alice would successfully verify the ownership of the suspected model by issuing prediction queries of key samples.

To evade the verification by the legitimate owners, Eve will attempt to build a detector to detect whether the queried sample is a clean one or a possible key sample \cite{HM18.0}. 
%According to the detection result, the queried sample will either be fed to the licensed model or be rejected without returning any results. In addition to the case of rejection, 
Once the detector decides the queried instance is a possible key sample, the stolen model would return a random label from its output space. However, all the existing watermark methods \cite{ABCPK18,ZGJWSHM18,RCK18.0,GP18} are susceptible to the above form of attack. The distribution of the key samples and the original samples varies greatly, once the model owner issues prediction of these key samples, Eve can easily detect the watermarks by utilizing a sample detector. Hence, in the following sections, a blind-watermark based IPP framework is presented that is aiming to defend against the evasion attack. We conduct extensive experiments, and the results demonstrate that our novel IPP framework can achieve state-of-art performances on undetectability against evasion attack.

\subsection{Legality: Fraudulent Claims of Ownership}
Suppose there is a counterfeiter, Oscar, who tries to illegally claim the ownership of the model. This behavior will not only infringe on Bob's interests but will even infringe on Alice's interests, which will directly lead to the invalidation of the IPP technology --- the model owner Alice is no longer the only one that can claim the ownership. Counterfeiters will try to make a set of key samples by himself, that is to say, design a set of fake samples that can induce the behavior of the licensed model to achieve the purpose of falsehood.

In the watermarking method proposed by Zhang et al. \cite{ZGJWSHM18}, the three types of perturbations superimposed to the ordinary samples is so obvious and striking that Oscar can easily superimpose the same perturbations to the ordinary samples to obtain the key samples. Due to the intrinsic generalization and memorization capabilities of deep neural networks, the newly generated samples can still be identified and responded with predefined labels \cite{ZGJWSHM18}. In the approaches of Adi et al. \cite{ABCPK18} and Rouhani et al. \cite{RCK18.0}, the space of abstract images and random images is so large, Oscar can easily obtain another set of images, for example, generated by computer script or genetic algorithms \cite{GP17}. Furthermore, the counterfeiters even can monitor and intercept key samples on the communication channel of the network. 

In this paper, we make an assumption that the hyper-parameters of the training setting, the parameters, and the architecture of encoder are confidential. Therefore it is impossible for an attacker to get the same watermark generation strategies. Further encouragingly, even we relax this key assumption, our IPP framework still achieves a remarkable performance on unforgeability against fraudulent claims of ownership.

\section{Watermarking Neural Network}
\begin{figure*}[htbp]
\centerline{\includegraphics[width=17cm, height=5.5cm]{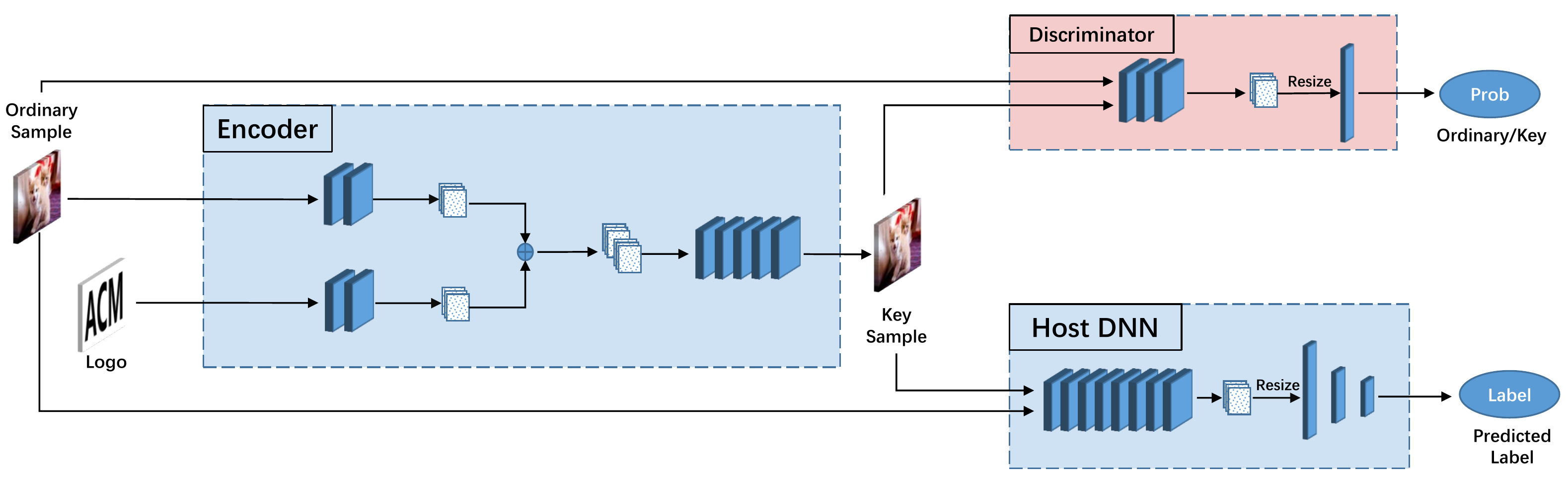}}
\caption{Workflow of our IPP framework}
\label{workflow}
\end{figure*}
In this section, we first briefly introduce the overview of embedding and verification of watermarking model. Hereafter, we introduced the details of the implementation of our IPP framework, including the objective functions and the watermarking algorithm.

\subsection{Tasks I: Embedding}
To protect the intellectual property of our model, we tried to watermark the model to leave the backdoor. In the embedding procedure, we need a set of labeled samples, which is also known as the key samples $x^{key}$, to be a watermark. The key sample $x^{key}$ is generated by an encoder $e$ which accepts the original samples $x$ and logo $l$ as inputs. In this paper, we hope that the distribution of the key sample $x^{key}$ is as close to that of the original samples $x$. To this end, we present a novel  generation algorithm $\mathcal{G}$ to achieve it:
\begin{align}
    &x^{key} = \mathcal{G}(e, x, l), \\
    &x^{key} \rightarrow  x \nonumber
\end{align}
Then we adopt an embedding algorithm $\mathcal{E}$ to embed a watermark to model $f$:
\begin{align}
    f_k = \mathcal{E}(f, x^{key}) 
\end{align}
The resultant model $f_k$ (i.e., the watermarked model) will predict a query of key sample $x^{key}$ to a pre-defined label $t^{key}$. Details of each algorithm are explained in the next section.

\subsection{Task II: Verification}
Consider a scenario that the model owner suspects that the model deployed remotely violates its copyright interest.
To confirm the ownership of the remote DNN, in this procedure, the model owner first prepares a set of key samples $\{ x_1^{key}, x_2^{key}, ... \}$ by a generation algorithm $\mathcal{G}$:
\begin{align}
    x^{key} = \mathcal{G}(e, x, l)
\end{align}
Then the model owner will issue a prediction query to the remote model $g$ with these key samples, obtain resulting predictions, and evaluate the accuracy of the resulting predictions over pre-defined labels.
\begin{align}
    acc_g = \mathcal{V}(g, x^{key}, t^{key})
\end{align}
If the $acc_g$ is a value close to 1, or $acc_g > \mathcal{T}_{acc}$ , where $\mathcal{T}_{acc}$ is a threshold parameter close to 1. Then the owner can verify the IP of a suspected model and claim the ownership of the remote model.

\subsection{Algorithm Pipeline} 
Figure \ref{workflow} shows the workflow of our IPP framework, which consists of three parts: encoder, discriminator and host DNN.

\textbf{Encoder:}
Here our encoder $e$ is essentially a lightweight autoencoder. The encoder accepts the part samples from the training dataset and the exclusive logo as inputs and attempts to output the key samples which are undistinguished from the ordinary samples. %It is unsupervised learning that learns how to compress and encode input data efficiently, and then learn how to reconstruct the data from a compression-encoded representation back to the representation of the original input as similar as possible. 
We typically denote the parameters of encoder as $\theta_e$, the exclusive logo as $l$, and try to obtain a function $\theta_e(x,l)=x^{key}$ where $x^{key} \to x$ by solving the optimization problem with a batch of $\{x_1, x_2, ..., x_m\}$:
\begin{equation}
\begin{aligned}
    \underset{\theta_e}{\operatorname{argmin}}
     \frac{1}{m} \sum_{i=1}^{m}(x_i-\theta_e(x_i,l) )^2
     \label{lemse}
\end{aligned}
\end{equation}
The above term is the reconstruction error for the encoder $e$.
%typically through a variant of stochastic gradient descent, yields the optimal parameters $\theta ^* $ of the encoder $E$.
%Where $m$ is the batch-size of input data. 

Due to the limited capacity of the encoder, it is impossible to achieve the goal of perfect reconstruction. Moreover, compared to perfect reconstruction, we hope that the distribution of key samples generated by the encoder only needs to be as close as that of ordinary samples, i.e., $x \approx x^{key}$, not $x = x^{key}$. The tiny difference between the key samples and the ordinary samples is exactly what we need --- the magnitude of fluctuation achieves a comparable trade-off between security and effectiveness, the smaller preserving better security against evasion attack and the larger providing better effectiveness of watermarking DNN.

From a mathematical perspective, in order to prove whether the objective function \ref{lemse} is exact to achieve the goal, i.e., $x \approx x^{key}$, not $x = x^{key}$, we denote the distribution of original samples from training dataset as $P_{data}(x)$, and the distribution of key samples produced by the encoder as $P_e (x^{key}; \theta_e)$. The objective function can be formalized as follows:
\begin{equation}
\begin{aligned}
    \underset{\theta_e}{\operatorname{argmax}}\prod_{i=1}^{m} {P}_{e}(x_i; \theta_e)
    =&\underset{\theta_e}{\operatorname{argmax}} \log \prod_{i=1}^{m} P_{e}(x_i; \theta_e)\\
    =&\underset{\theta_e}{\operatorname{argmax}} \sum_{i=1}^{m} \log P_{e}(x_i;\theta_e)\\ 
    =& \underset{\theta_e}{\operatorname{argmax}} \mathbb{E}_{x \sim P_{\text {data}}}\left[\log P_{e}(x; {\theta_e})\right]\\
    =&\underset{\theta_e}{\operatorname{argmax}} \int_{x} P_{data}(x) \log P_{e}(x; \theta_e)dx\\
    &-\int_{x} P_{data}(x) \log P_{data}(x)dx\\
    =&\underset{\theta_e}{\operatorname{argmin}} KL\left(P_{data}(x)\| P_{e}(x; \theta_e)\right)
\end{aligned}
\label{KL}
\end{equation}
where KL is the Kullback-Leibler divergence. The above objective function is essential to minimize the Kullback-Leibler divergence which is a measure of how one probability distribution diverges from another. Further derivation:
\begin{align}
    KL(P_{data}\|P_{e})= -H(P_{data})+H(P_{data}, P_e)
\label{inen}
\end{align}
The former of equation \ref{inen} represents the information entropy of $P_{data}$, the latter is the cross entropy of $P_{data}$ and $P_{e}$. That is to say, minimizing the KL divergence is equivalent to minimizing the cross entropy. At the same time, the objective function \ref{lemse} is actually the cross entropy of the empirical distribution and the gaussian model \cite{GBC16}, while we cannot determine which distribution $P_{data}$ and $P_{e}$ obey. In order to solve the objective function \ref{KL}, we adopt the negative sampling approach \cite{MSCCD13}.
Furthermore, the equation \ref{lemse} only punishes the larger error of the corresponding pixels of the two images, and ignores the underlying structure of the image. So we introduce the structural similarity index (SSIM) \cite{ssim}, and the objective function can be formalized as follows:
\begin{align}
     %SSIM(I, I') &= \frac{(2\mu_I\mu_{I'}+c_1)(2\sigma_{II'}+c_2)}{({\mu_I}^2+{\mu_{I'}}^2+c_1)({\sigma_{I}}^2+{\sigma_{I'}^2}+c_2)}\nonumber\\
     \underset{\theta_e}{\operatorname{argmin}}
     \frac{1}{m}& \sum_{i=1}^{m}(1-SSIM(x_i,\theta_e(x_i,l)))\label{1ssim}
\end{align}
%In view of the above, we have introduced the adversarial training to obtain the optimal $\theta ^*$ by minimizing the KL divergence.

\textbf{Discriminator:} The negative sampling approach targets a different objective than the original function \ref{KL}. 
%As in the actual situation, we are unable to determine what distribution the $P_{data} (x)$, $P_E (x^{key}; \theta)$ obeys. In order to minimize the KL divergence between $P_{data}(x)$ and $P_{E}(x; \theta)$,
We typically interpret this objective as a binary classification problem. The discriminator efforts to determine whether the samples are synthesized or extracted from the ordinary samples and the encoder converges to capture the distribution of given samples. %The final result of the game is that the encoder can generate key samples which are enough to mix the false with the genuine. For the discriminator, it is difficult to distinguish whether the generated samples are real or not, which indicates the KL divergence between $P_{data}(x)$ and $P_{e}(x; \theta_e)$ has been minimized. 
The keypoint of the discriminator is essentially the theory proposed in generative adversarial networks \cite{IJMBDSAY}, which is well-designed to minimize the KL divergence exactly. The discriminator also acts as a detector to detect if input data is generated by the encoder.

We denote the distribution of training dataset $P_{data}(x)$ as a positive sample, that of the key samples $P_{e}(x^{key}; \theta_e)$ as a negative sample. In order to speed up the learning process, we adopt $x$ and $x^{key}$ to present the distribution $P$ which is unable to obtain.
The discriminator $d$ accepts the original samples $x$ or key samples $x^{key}$ as inputs, outputting a binary classification probability to indicate whether the input data comes from the ground truth. The discriminator outputs a conditional probability $p(\Theta|\chi; \theta_{d})$ ($\chi\in \{x,x^{key}\}$) which is modeled as logistic regression:
\begin{equation}
p(\Theta  |\chi; \theta_d)=\frac{1}{1+e^{-\theta_d(\chi)}}
\end{equation}
%By maximizing the condition probability $p(\Theta|\chi; \theta_{d})$, we can obtain the optimal $\theta_d$ which indicates the discriminator can successfully determine whether the input samples are original or not.
where we use $\theta_d$ to represent the discriminator and adopt a random variable $\Theta$ to denote the binary answer: $\Theta =1$ if the input sample is the origin, and $\Theta =0$ otherwise. \textbf{The objective function ($O_d$) of the discriminator is denoted by}:
\begin{equation}
 \begin{aligned}
    \underset{\theta_d}{\operatorname{argmin}}-\frac{1}{m}\sum_{i=1}^m\Bigg(\Theta\log \frac{1}{1+e^{-\theta_d(\chi_i)}}+\\(1-\Theta)\log (1-\frac{1}{1+e^{-\theta_d(\chi_i)}})\Bigg)
\label{logistic_d}
\end{aligned}   
\end{equation}
After the optimal $\theta_d$ is given, we simultaneously train the encoder $e$ to maximize the probability of $d$ making a mistake. Then, the new objective function of encoder $e$ is:
\begin{align}
\label{e_d}
    %\underset{\theta_e}{\operatorname{argmax}}\sum_{\chi\in \{x_1^{key},x_2^{key},...,x_m^{key}\}}^m\log \frac{1}{1+e^{-\theta_d(\chi)}}=\nonumber\\
    \underset{\theta_e}{\operatorname{argmin}}-\frac{1}{m}\sum_{i=1}^m\log \frac{1}{1+e^{-\theta_d(\theta_e(x_i,l))}}
\end{align} 
We train the discriminator $d$ and the encoder $e$ iteratively to obtain the final optimal $\theta_d^*$ and $\theta_e^*$. Competition in this procedure derives both teams to improve their ability until the $P_e(x^{key};\theta_e)$ is indistinguishable from the genuine $P_{data}(x)$. Note that, both of the objective function \ref{lemse}, \ref{1ssim} and \ref{e_d} regularize the encoder $e$ by encouraging the distribution $P_{e}(x^{key}; \theta_e)$ to match the given distribution $P_{data}(x)$. Next, we introduce how to induce the host model to mispredict a key sample to a pre-defined label.

\textbf{Host DNN:} Due to the capacity limitations of the encoder $e$, the distribution of key samples generated from encoder can only be "close" to that of ordinary samples. That is, there must be a "some" difference between $x^{key}$ and $x$. Hence, we utilize this difference to induce the host DNN ($\theta_h$) to correctly identify $x^{key}$ to a pre-defined label under the premise of correctly identifying $x$. In the image classification task, it is common to %adopt the cross-entropy to denote the distance between the two probability distributions. However, the output vector $\theta_h(x)$ of the host model $h$ is not necessarily a probability distribution. DNN for classification typically 
employ the softmax function at the final layer to obtain the probability vector:
\begin{align}
\label{mind}
    softmax(\theta_h(x))_i = \frac{e^{\theta_h(x)_i}}{\sum_j^n e^{\theta_h(x)_j}}
\end{align}
Where $j$ denotes the $j$-th element of the output vector of $n$-class. Given the original sample $x$ with normal label $t$, and key sample $x^{key}$ with a pre-defined label $t^{key}$, 
\iffalse
we obtain two objective functions of $h$:
\begin{align}
    \underset{\theta_h}{\operatorname{argmin}}-\sum_{\chi\in\{x_1, x_2, ...,x_m\}}^m\left(\log\frac{e^{\theta_h(\chi)_{t}}}{\sum_j^n e^{\theta_h(\chi)_j}}\right)
    \\
    \underset{\theta_h}{\operatorname{argmax}}\sum_{\chi\in\{x_1^{key}, x_2^{key}, ...,x_m^{key}\}}^m\left(\log\frac{e^{\theta_h(\chi)_{t^{key}}}}{\sum_j^n e^{\theta_h(\chi)_j}}\right)\label{h_k}
\end{align}
\fi
we briefly define $\mathcal{D} = \{(x_1, t_1), ..., (x_1^{key}, t_1^{key}), ...\}$, then \textbf{the objective function ($O_h$) of the host model $h$ is denoted by}:
\begin{align}
    \underset{\theta_h}{\operatorname{argmin}}-\frac{1}{m}\sum_{(\chi,\tau)\in\mathcal{D}}^m\left(\log\frac{e^{\theta_h(\chi)_{\tau}}}{\sum_j^n e^{\theta_h(\chi)_j}}\right)
    \label{h_loss}
\end{align}
Note that, the equation \ref{h_loss} of which $x^{key}=\theta_e(x,l)$ with label $t^{key}$ can also be formalized to update $\theta_e$ as follows:
\begin{align}
\label{e_h}
    \underset{\theta_e}{\operatorname{argmin}}-\frac{1}{m}\sum_{i=1}^m\left(\log\frac{e^{\theta_h(\theta_e(x_i;l))_{t_{i}^{key}}}}{\sum_j^n e^{\theta_h(\theta_e(x_i;l))_j}}\right)
\end{align}
By adding all the equations \ref{lemse}, \ref{1ssim}, \ref{e_d} and \ref{e_h} together, we have \textbf{the following objective function ($O_e$) for encoder $e$}:
%\begin{equation}
\begin{align}
 &\underset{\theta_e}{\operatorname{argmin}}\frac{1}{m}\sum_{i=1}^m \Bigg(\alpha (x_i-\theta_e(x_i,l))^2 +\beta(1-SSIM(x_i,\theta_e(x_i,l))\nonumber\\&+\gamma (-\log \frac{1}{1+e^{-\theta_d(\theta_e(x_i,l))}}) +
\delta (-\log\frac{e^{\theta_h(\theta_e(x_i;l))_{t_{i}^{key}}}}{\sum_j^n e^{\theta_h(\theta_e(x_i;l))_j}} )\Bigg)
\end{align}
%\end{equation}
where $\alpha , \beta, \gamma, \delta>0$ are the hyper-parameters to trade-off between four parts. The 4th term regularizes the encoder $e$ by encouraging the generated key samples misclassified to pre-defined labels more easily. 
The global watermarking algorithm is as follow:
\begin{algorithm}  
    \caption{Minibatch gradient descent training of watermarking algorithm.}
    \label{alg:watermarking}  
    \KwIn{training set $D$, logo $l$, hyper-parameters $\alpha,\beta,\gamma,\delta$, minibatchsize $m,n$, sampling number $k_{1,2}$\;}
    \KwOut{$\theta_e,\theta_h,\theta_d$\;}  
    Initialize original samples $D'$ sampled from $D$ randomly\;
    For convenience, objective function is denoted by $O_{e,d,h}$\; %\varphi_n\bm{W}\vec{y}_n  
    \For{number of training epochs}
    {
     \For{$i=1;i \leq k_1;i++$ }{
     sample minibatch of m original samples from $D'$\;
     update $\theta_d$ by descending its adam gradient:
     $\nabla_{\theta_{d}}O_d$\;
     update $\theta_e$ by descending its adam gradient:
     $\nabla_{\theta_{e}}O_e$\;}
     \For{$i=1;i \leq k_2;i++$}{
     sample minibatch of n sample from D\;
     merge minibatch of $m+n$\;
     update $\theta_h$ by descending its stochastic gradient:
     $\nabla_{\theta_{h}}O_h$\;
     }
     }
    return $\theta_e,\theta_h, \theta_d$\;  
    %\end{algorithmic}  
\end{algorithm} 
\begin{table*}[htbp]
\centering
\caption{Details of the DNNs and datasets used to evaluate our IPP framework}
\begin{threeparttable}
\setlength{\tabcolsep}{5mm}{
\begin{tabular}{l|c|c|c|c}  
\toprule
\multirow{2}*{Dataset}&Dataset &Host DNN  & Host DNN      & Test   \\ 
&Description&Architecture&Description&Acc.\tnote{\#}\\
\midrule
\multirow{3}*{MNIST\cite{MNIST}}&\multirow{3}*{\shortstack{Hand-written\\ digits}}&LeNet-1\cite{LBBHn} & \multirow{3}*{\shortstack{A classic CNN Architecture}}  & 98.73\%   \\

&&LeNet-3\cite{LBBHn}& &  98.86\%   \\

&&LeNet-5\cite{LBBHn}& &   98.92\%   \\
\midrule

\multirow{12}*{CIFAR-10\cite{CIFAR}}&  \multirow{12}*{\shortstack{A collection of\\ General images}} 
     &VGG-11\cite{SZ14.0} & \multirow{4}*{\shortstack{Very Deep \\Convolutional Networks \\for Large-Scale \\Image Recognition}} &  91.40\%  \\

&&VGG-13\cite{SZ14.0} & &  93.47\%   \\

&&VGG-16\cite{SZ14.0} & &  93.08\%    \\

&&VGG-19\cite{SZ14.0} & &  92.86\%    \\
\cmidrule{3-5}

&&ResNet-18\cite{HZRS16} & \multirow{3}*{\shortstack{Deep residual \\learning for \\image recognition}} &  94.62\%    \\

&&ResNet-34\cite{HZRS16} & &  94.80\%   \\

%&&ResNet-50\cite{HZRS16} & &  93.16\%    \\

&&ResNet-101\cite{HZRS16} & & 93.97\%    \\

\cmidrule{3-5}
&&PreActResNet-18\cite{HZRS162} &\multirow{2}*{\shortstack{Identity Mappings in \\Deep Residual Networks}} &  94.43\%    \\
&&PreActResNet-34\cite{HZRS162} & &  94.95\%    \\

\cmidrule{3-5}
&&GoogleNet\cite{SLJSRAEVR15} &Going Deeper with Convolutions &  94.38\%    \\

\cmidrule{3-5}
&&DPN-26\cite{CLXJYF17} &Dual Path Networks &  94.53\% \\

\cmidrule{3-5}
&&MobileNetV2\cite{HZCKWWAA17.0} & Efficient CNN for Mobile App&  91.35\%  \\

\bottomrule
\end{tabular}}
\begin{tablenotes}
        \footnotesize
        \item[\#] The accuracy is obtained from regular test set in unwatermarked setting.
       
      \end{tablenotes}
\end{threeparttable}
\label{summary}
\end{table*}
\begin{figure*}[t]
\centering
\subfigure{\includegraphics[width=0.1\linewidth]{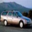}}
\subfigure{\includegraphics[width=0.1\linewidth]{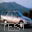}}
%\subfigure[]{\includegraphics[width=0.09\linewidth]{figure/car/t_b.png}}
\subfigure{\includegraphics[width=0.1\linewidth]{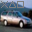}}
%\subfigure[]{\includegraphics[width=0.09\linewidth]{figure/car/g_b.png}}
\subfigure{\includegraphics[width=0.1\linewidth]{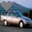}}
\subfigure{\includegraphics[width=0.1\linewidth]{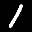}}
\subfigure{\includegraphics[width=0.1\linewidth]{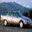}}
\subfigure{\includegraphics[width=0.1\linewidth]{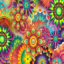}}
\subfigure{\includegraphics[width=0.1\linewidth]{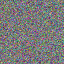}}
\subfigure{\includegraphics[width=0.1\linewidth]{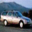}}

\subfigure[(a) Original]{\includegraphics[width=0.1\linewidth]{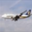}}
\subfigure[(b) "test" \cite{ZGJWSHM18}]{\includegraphics[width=0.1\linewidth]{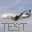}}
%\subfigure[]{\includegraphics[width=0.09\linewidth]{figure/car/t_b.png}}
\subfigure[(c) "symbol" \cite{ZGJWSHM18}]{\includegraphics[width=0.1\linewidth]{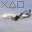}}
%\subfigure[]{\includegraphics[width=0.09\linewidth]{figure/car/g_b.png}}
\subfigure[(d) "heart" \cite{ZGJWSHM18}]{\includegraphics[width=0.1\linewidth]{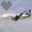}}
\subfigure[(e)"unrelated" \cite{ZGJWSHM18}]{\includegraphics[width=0.1\linewidth]{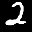}}
\subfigure[(f) "mask" \cite{GP18}]{\includegraphics[width=0.1\linewidth]{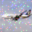}}
\subfigure[(g) "abstract" \cite{ABCPK18}]{\includegraphics[width=0.1\linewidth]{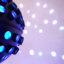}}
\subfigure[(h) "random" \cite{RCK18.0}]{\includegraphics[width=0.1\linewidth]{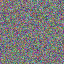}}
\subfigure[(j) Ours]{\includegraphics[width=0.1\linewidth]{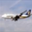}}
\caption{The examples of key samples of existing watermark methods and our framework}
\label{compare}
\end{figure*}
\section{Implementation}
\subsection{Datasets and DNNs}
As a proof-of-concept, we adopt two benchmark datasets with different types of data---MNIST and CIFAR-10---and implement our IPP framework on a total of fifteen host DNNs. We provide a summary of the two datasets and the corresponding DNNs in Table \ref{summary}. %The detailed architectures of encoder and discriminator can be found in Appendix \ref{appendix}.
%will be made available soon\textsuperscript{\ref{code}}.

\textbf{MNIST} \cite{MNIST} is a large handwritten digital dataset containing $28 \times 28$ pixel images with class labels from 0 to 9. The dataset consists of 60,000 training samples and 10,000 test samples. Each pixel value is within a grayscale between 0 and 255.

\textbf{CIFAR-10} \cite{CIFAR} is labeled subsets of the 80 million tiny color images dataset, consisting of 50,000 training images (10 classes, 5,000 images per class) and 10,000 test images (10 classes, 1000 images per class). All the images are normalized and centered in a fixed-size image of $32 \times 32$ pixels.

To evaluate our blind-watermark based IPP framework, we use 1\%  of total training samples for the key sample generation, and for each image, we randomly select a target label. The encoder and discriminator are trained iteratively using Adam algorithm \cite{KB14.0} ($\beta_1=0.5, \beta_2=0.999$) with a mini batch-size of 20 and a fixed learning rate of 0.001. The host DNNs are simultaneously trained using stochastic gradient descent \cite{HM51.0} with a batch-size of 120 (100 original samples and 20 key samples) and a declining learning rate of 0.1, which is decayed by 0.1 per 40 epochs. we perform grid search to find the optimal hyper-parameters $\alpha = 3, \beta = 5, \gamma = 0.1, \delta = 0.01$ and train a prototype of our framework for 100 epochs with $k_{1,2}=1$. The detailed architectures of encoder and discriminator can be found in Appendix \ref{appendix_arch}. All our experiments are implemented in Python with Pytorch \cite{PyTorch}, on a Ubuntu 18.04 server with a Tesla K80 GPU card. %For reproducibility purposes, our code is available at \url{https://github.com/zhenglisec/Blind-Watermark-for-DNN}

%We implement all the experiments\footnote{Source code is available at \url{https://github.com/zhenglisec/Blind-Watermark-for-DNN}} by Pytorch \cite{PyTorch}, on a Ubuntu 18.04 server with a Tesla K80 GPU card.
\begin{figure*}[htbp]
\centerline{\includegraphics[width=17cm, height=5.5cm]{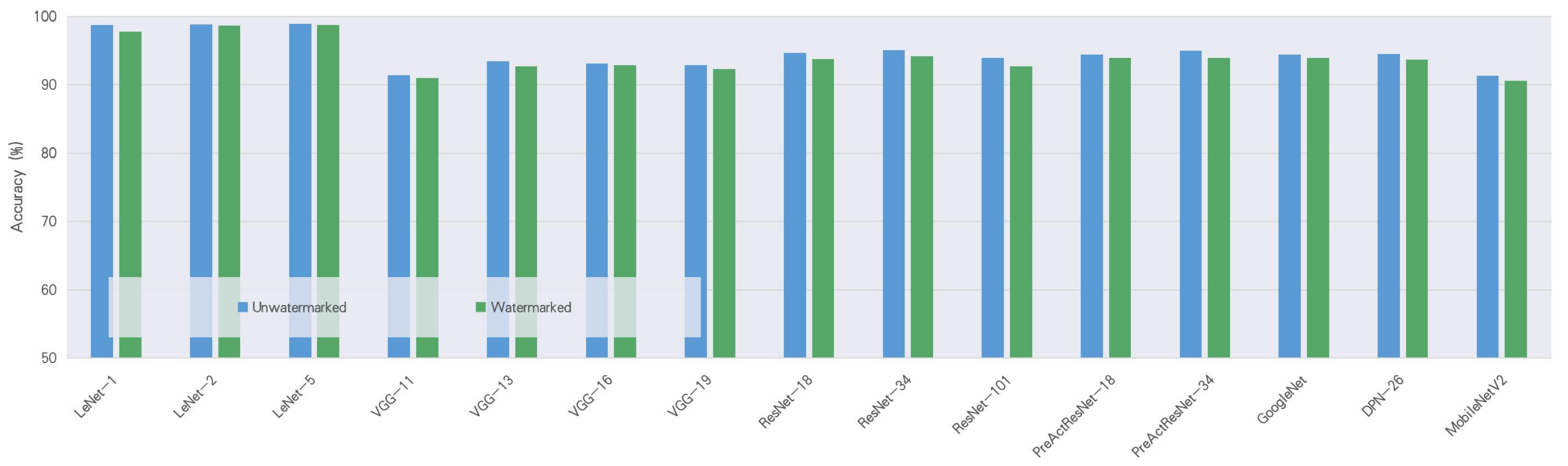}}
\caption{Accuracy of different models on regular test set}
\label{fidelity}
\end{figure*}
\subsection{Results}
The training of our IPP framework has been successfully implemented, and we compare the key samples generated by our framework with the existing methods. Figure \ref{compare} shows examples of key samples created from CIFAR-10 dataset. Figure \ref{compare} (a) are two examples of original images; Figure \ref{compare} (b)-(h) are key samples generated by methods proposed in \cite{ABCPK18,GP18,RCK18.0,ZGJWSHM18}. Figure \ref{compare} (j) are two key samples generated by our blind-watermark based IPP framework. As we can see, the examples of ours are so similar to the original samples that the differences between them are too tiny to be seen by humans. In contrast, the examples of other existing methods are visible and striking, which indicates the distribution of the features of them is distant from the feature distribution of the training samples. In next section, we conduct extensive empirical validations to show that our IPP framework satisfies multiple requirements.

\section{Evaluation}

We analyze the performance of our IPP framework by measuring the following criteria: \textbf{fidelity}, the side effect made to the primary classification task; \textbf{effectiveness and  integrity}, whether it can successfully verify the ownership of the host DNN; \textbf{security}, the ability of defending against evasion attack; \textbf{legality}, the ability of anti-counterfeiting; \textbf{feasibility}, The ability to resist model modifications and whether it explicitly associates the model with the identity of the actual creator.
% "proof-of-origin" problem. It is different to the "proof-of-authorship" during actual model usage\cite{guo2018watermarking}.

\subsection{Fidelity}

Fidelity requires our IPP framework to watermark a host model without significant side effects on the primary task. Ideally, a well-watermarked model should be as accurate as an unwatermarked model. To measure the side effects on the primary task, we implemented a comparative evaluation of the accuracy between the clean model and watermarked model. As depicted in Figure \ref{fidelity}, all the evaluated models are trained on the test set in two different settings: unwatermarked setting and watermarked setting. We first train a model without watermark embedding and evaluate it on the test set that it has not seen before. Then we implement our framework to watermark the same model and evaluate it on the test set.

The results expressly demonstrate that all the watermarked models still have the same level of accuracy as the unwatermarked model. The accuracy drops by up to 0.66\% on average. In the best case, we achieve a drop of only 0.14\%. That is to say, the side effects caused by our IPP framework are entirely within the acceptable performance variation of the model and has no significant impact on the primary task. Thus our framework meets the fidelity requirement.
\begin{figure*}[htbp]
\centerline{\includegraphics[width=17cm, height=5.5cm]{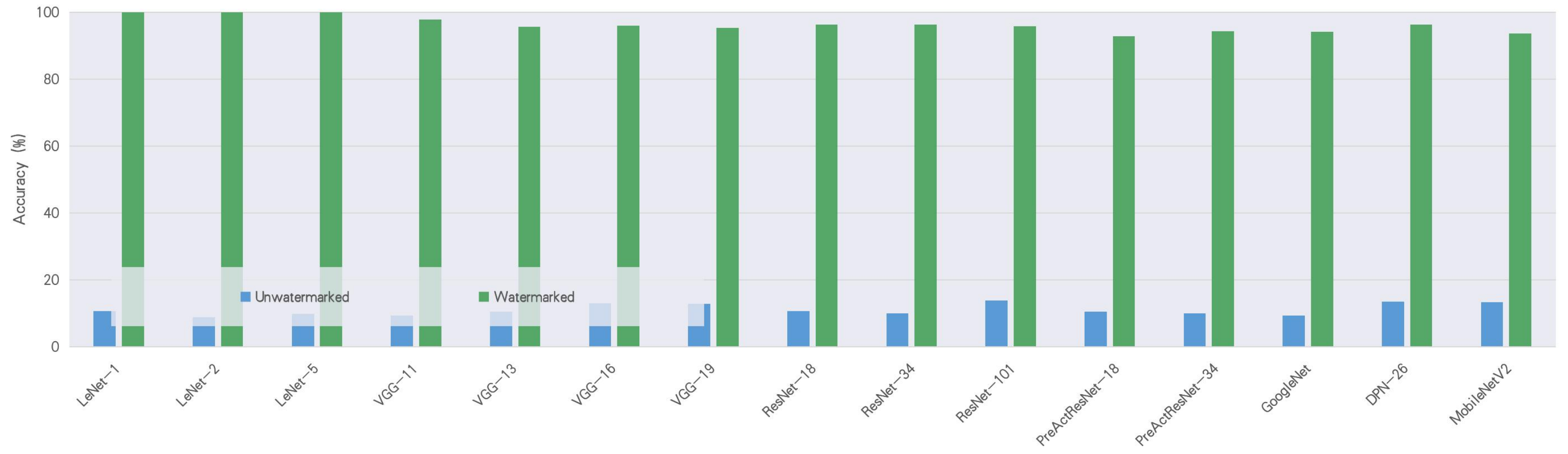}
}
\caption{Accuracy of different models on our key samples}
\label{effectiveness}
\end{figure*}

\subsection{Effectiveness and Integrity}
The purpose of effectiveness is to measure whether we can successfully verify the copyright of the target DNN model under the protection of our IPP framework. Ideally, a well-watermarked model should identify key samples and predict them to the pre-defined labels with high accuracy. The integrity requires that our IPP framework shall not falsely claim the authorship of unwatermarked models. To measure the effectiveness and integrity, we implement another comparative evaluation of the accuracy between the unwatermarked model and the watermarked model. We typically denote the forward inference function of the unwatermarked model by $\theta_{uw}$ and that of the watermarked model by $\theta_w$.  Only if $\theta_{uw}(x^{key})\neq t^{key}$ and $\theta_w(x^{key})==t^{key}$, we confirm that our IPP framework can successfully verify the ownership. 
We issue prediction queries of key samples and tests whether the model returns correct labels specified by key samples. 

Figure \ref{effectiveness} shows the accuracy of the different models implemented in our evaluation. We use $1\%$ of dataset to test the accuracy. The "unwatermarked" shows the accuracy of key samples that are not induced into the unwatermarked model. All the unwatermarked models achieve an accuracy of 9\%---13\%, a totally random guess. In contrast, the "watermarked" encouragingly shows the accuracy of key samples exceeds over 90\%. In the best case, the watermarked model even achieve an accuracy of 100\% on the key samples. The results show that our framework can successfully verify the ownership without falsely claiming the authorship of unwatermarked models. Thus the effectiveness and the integrity requirements are met.

%Security requires that the key samples generated by our IPP framework shall be imperceptible and undetectable to prevent identification or detection by the unauthorized service provider.

\subsection{Security}
Security requires that the verification process is imperceptible and undetectable, which can resist identification and detection by unauthorized service providers. In this section, we introduce a novel watermark invalidation method, evasion attack \cite{HM18.0}, and demonstrate that an attacker can evade the verification by the model owners. % under the protection of existing watermarking techniques. The attacker would build a detector based on neural networks, and train the detector as a binary classifier. The licensed model or stolen model will be deployed behind the detector. And the detector will attempt to distinguish whether the sample of the current query is a key sample or an ordinary sample. Relying on the detecting result, the current query either be forward the illegally used model or be rejected. In general, it will return an arbitrary label from the output space of the illegally used model. 
Considering that the premise of illegal service provision is due to the limitation of data resources or computing power, we make an assumption that an attacker can access to 10\% of total training samples for detector build-up. %Moreover, if the attacker can access to a massive collection of training samples, the attacker can use the massive samples to train a sufficiently high-performance model. That is, there is no motivation for the attacker to steal the licensed model. Therefore, we suppose that the attacker has a small number of samples. 
Essentially, the binary classification is supervised learning. They can generate the same amount of key samples according to the proposed methods in \cite{GP18,ABCPK18,ZGJWSHM18,RCK18.0} and our framework, then get the training set for detector build-up. We typically design a simple detector based on 3 fully-connected layers, which are composed of 512, 256 and 1 neural unit followed by sigmoid activation. We use $\theta_d$ to represent the parameters of the detector, %and adopt a random variable $\bigtriangleup$ to denote the binary answer: $\bigtriangleup=1$ if the current queried sample is a possible key sample, and $\bigtriangleup=0$ otherwise. 
%and the detector outputs a conditional probability $p(\bigtriangleup|\chi; \theta_{d})$ ($\chi\in \{x,x^{key}\}$) which is modeled as logistic regression. 
and the objective function is denoted by:
\begin{equation}
    \begin{aligned}
    \label{logistic}
    \underset{\theta_d}{\operatorname{argmin}}-\frac{1}{m}\sum_{i=1}^{m}\log \frac{1}{1+e^{-\theta_d(x_i)}}-\\\frac{1}{m}\sum_{i=1}^{m}\log (1-\frac{1}{1+e^{-\theta_d(x_i^{key})}})
\end{aligned}
\end{equation}
\begin{figure}[t]
\centerline{\includegraphics[width=6cm, height=5cm]{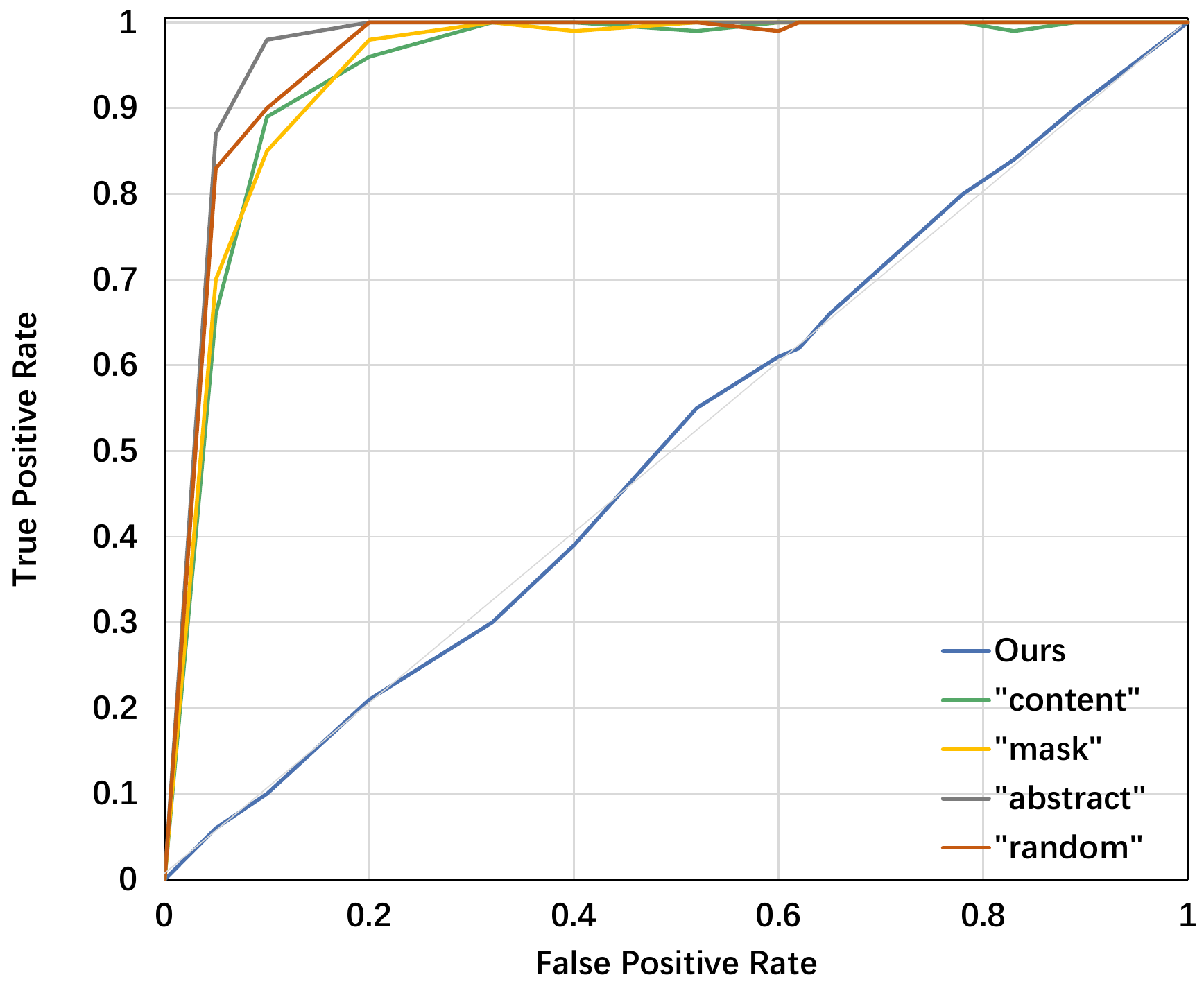}}
\caption{The receiver operating characteristic (ROC) curve produced by the detector based on 3 linear layers}
\label{roc3}
\end{figure}
We apply stochastic gradient descent with a fixed learning rate of 0.01 for 20 epochs to solve the objective function \ref{logistic}. We adopt the ROC (receiver operating characteristic), which reports the relation between true positive rate and false negative rate over multiple thresholds, as the evaluation metric.
\begin{figure}[htbp]
\centerline{\includegraphics[width=6cm, height=5cm]{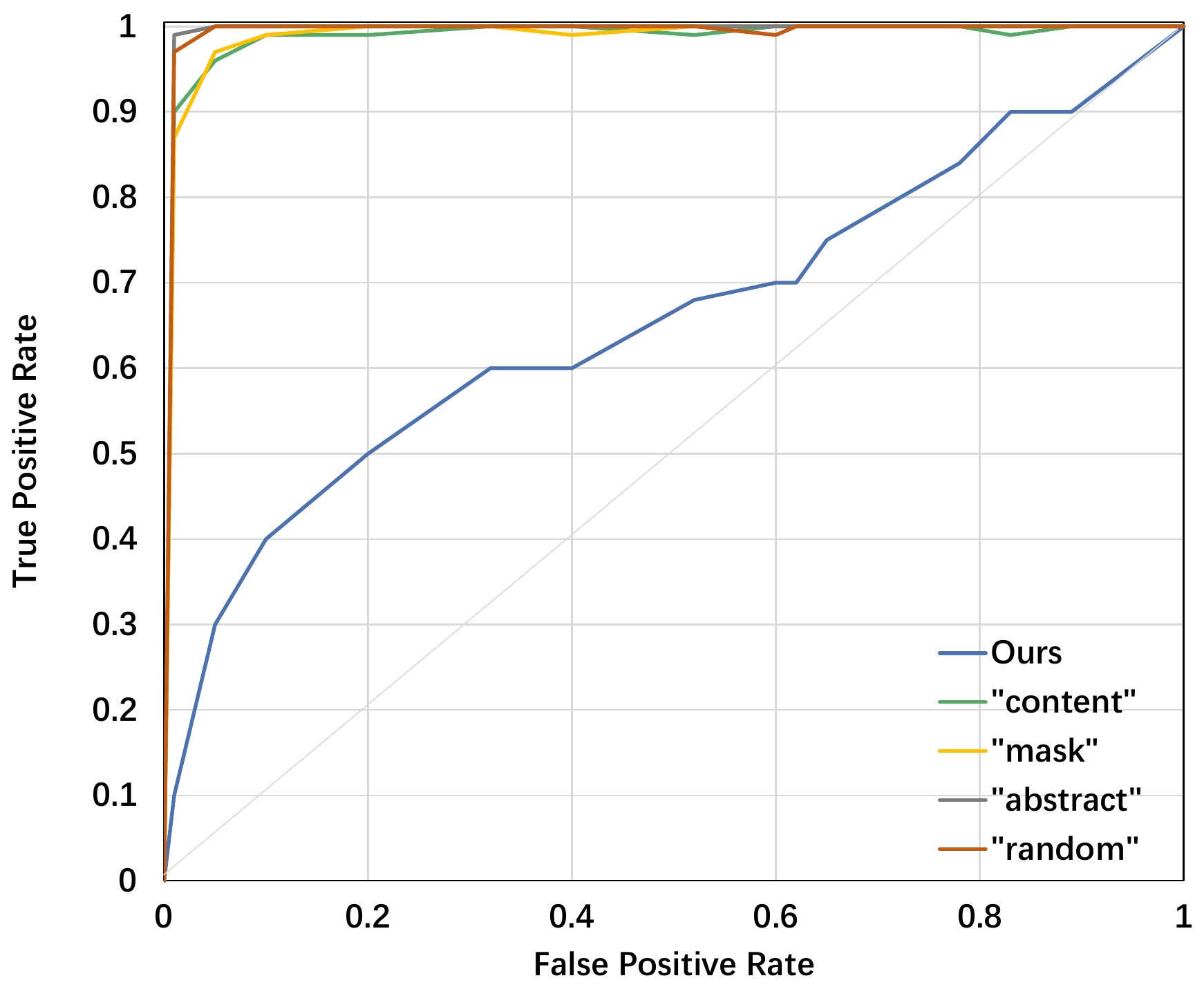}}
\caption{The receiver operating characteristic (ROC) curve produced by the detector based on ResNet-18}
\label{roc18}
\end{figure}
Figure \ref{roc3} presents the trained 3-layers detector's performance over the queried samples. The ROC curve of our IPP framework demonstrates that the performance of the detector is close to random guessing with an AUC (area under the ROC curve) of 0.5. In contrast, the ROC curve of previous methods demonstrates that the detector has a performance on AUC of well above 0.94, which indicates that the unauthorized service provider is enough to evade the verification. 

A further assumption is that the attacker tries to build a more powerful detector based on the weights transferred from the stolen model. In addition to the last several fully connected layers, most of the current classification models are playing the role of feature extractors. Therefore, we utilize the former layers of ResNet-18 as a feature extractor, which is then followed by a fully-connected layer with one output. We trained the detector in the same setting as the 3-layer detector and report the results in Figure \ref{roc18}. As we can see, the detection performance has indeed increased. The ROC curves of the existing methods show that the key samples used in their methods are more easily detected with an AUC of 0.98. Encouragingly, the ROC curve of our framework demonstrates that the detector is only slightly more effective than random guessing with an AUC of 0.65. This result convincingly shows that our IPP framework can achieve remarkable performances on undetectability against evasion attack.

In addition to evasion attack, we also consider another type of attack: removing the backdoor-based watermark. For example, an attacker can fine-tune the stolen model to achieve the purpose of removing the watermark. More often in practice, it is common to fine-tune the existing state-of-the-art models on new insufficient datasets to achieve higher performance or to implement new tasks. As for an authorized user, fine-tuning the model does not mean that he wants to launch this type of attack. Hence, we discuss the behavior of the fine-tuning model in detail in section \ref{robust}, which is regarded as a test of the robustness of our scheme.
\subsection{Legality}
Here, we consider an attack scenario in which the counterfeiter knows that the model purchased by Bob is watermarked and attempts to claim ownership of the model illegally. This behavior will not only infringe on Bob's interests but will even infringe on Alice's benefits, which will directly lead to the invalidation of the IPP technology --- the model owner Alice is no longer the only one that can claim the ownership. The counterfeiter attempts to designing a set of fake samples, which can induce the abnormal behavior of the licensed model. Therefore, the goal of legality is to resist the fraudulent claims of ownership by adversaries. In this paper, we study two different types of fraudulent claims of ownership.
\begin{figure}[t]
\centering
\subfigure{\includegraphics[width=0.22\linewidth]{figure/car/cover1.png}}
\subfigure{\includegraphics[width=0.22\linewidth]{figure/car/ours.png}}
\subfigure{\includegraphics[width=0.22\linewidth]{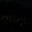}}
\subfigure{\includegraphics[width=0.22\linewidth]{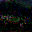}}

\subfigure[Original]{\includegraphics[width=0.22\linewidth]{figure/airplane/ours.png}}
\subfigure[Ours]{\includegraphics[width=0.22\linewidth]{figure/airplane/ours.png}}
\subfigure[Difference$\times$1]{\includegraphics[width=0.22\linewidth]{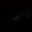}}
\subfigure[Difference$\times$5]{\includegraphics[width=0.22\linewidth]{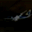}}
\caption{The examples of difference image}
\label{DF}
\end{figure}

\textbf{What if the ordinary samples and the key samples became accessible?} Although the ROC curve of our framework demonstrates that the detector based on ResNet-18 is only slightly more effective than random guessing with an AUC of 0.65, the counterfeiter can actually detect a small number of key samples.
%As we can see in the subsection[security],  the detector based on a neural network can successfully determine whether the queried samples are key samples or extracted from the ordinary samples against existing watermarking methods. Even though our framework, a detector based on the feature extractor of VGG-11, has an AUC performance of 0.6, that is, the counterfeiter can actually obtain a small number of key samples. 
Therefore, we assume that what if the original and key samples became accessible by intercepting the communication channel? What could then be ascertained about the intercepted samples? In Figure \ref{compare} (b)---(d), the features of superposed images are so prominent and striking that the counterfeiter can easily generate a set of fake samples by adding them to other original samples. Therefore, we apply the same logo "TEST" to other samples to generate a set of new key samples, and then issue prediction queries of the new key samples to the watermarked model, obtained an average accuracy higher than 91\%.
%In Table \ref{DF}, we showed the difference images between the ordinary samples and the key samples. As we can see, the features of the difference images represented by subtracting ordinary samples from their key samples are so prominent and striking that the counterfeiter can easily to generate a set of fake samples by adding them to other ordinary samples. As shown in Table \ref{DF}, we apply the same logo "TEST" to other samples to generate a new key samples, and then queried them to watermarked model, which still can mispredict them to a pre-defined label with an accuracy of above 90\%. 
In contrast, as depicted in Figure \ref{DF}, the features of the difference images from our framework are too subtle for the human to observe distortion, and we magnify all difference images by five times. It can be found that the distortion mode made upon each original sample is unique, and the distribution of distortions is related to the properties (e.g., complexity and texture) of the original samples. This result indicates that it is impossible to design a set of fake samples by superposing the difference images to other new original samples.

\begin{figure}[htbp]
\centerline{\includegraphics[width=6.5cm, height=5.6cm]{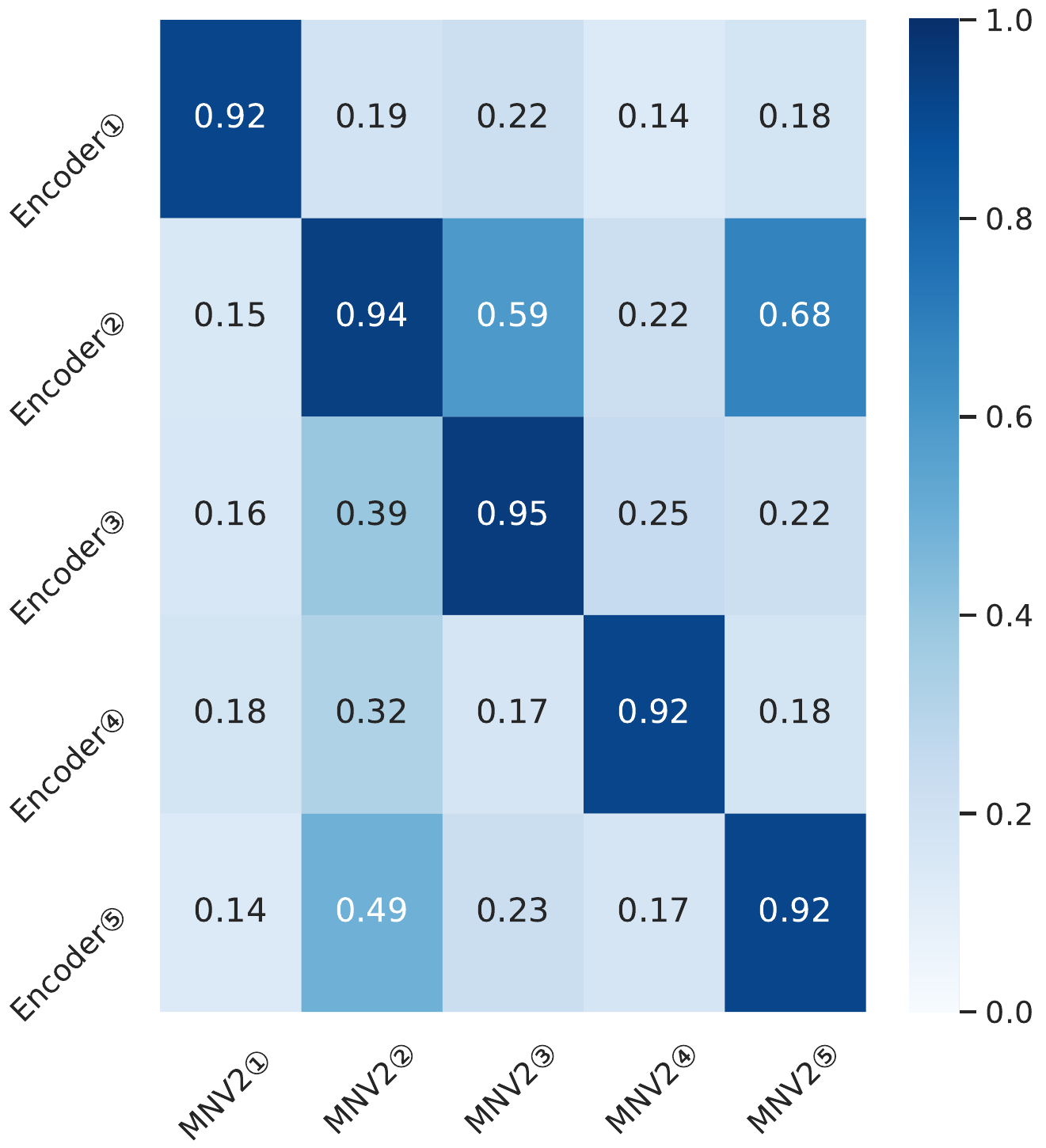}}
\caption{The performance (accuracy) of our key samples transferring attack}
\label{heatmap}
\end{figure}

\textbf{What if the encoder was leaked?} In most cases, it can safely be assumed that access to the learned encoder directly is impossible for an attacker. However, what if the attacker trained a "same" encoder by using the same architecture, dataset, and hyper-parameters?   To test this attack, we implement our IPP framework to watermark MobileNetV2 for five times with different seeds, then we get five identical pairs of encoder and
MobileNetV2 (MNV2), which are numbered \textcircled{1}, \textcircled{2}, \textcircled{3}, \textcircled{4} and \textcircled{5}. Figure \ref{heatmap} depicts the key samples transferring attack's performance. The x-axis represents the host model being attacked, and the y-axis represents the key samples generated by the trained encoder. Concretely, we issue queries to each MobileNetV2 with the key samples generated by each encoder to evaluate the accuracy of predicting the pre-defined labels. As we can see, the high accuracy of the attack results is listed at the diagonal of Figure \ref{heatmap}, which shows that the key samples can only induce their corresponding model to pre-defined labels. The reason why it can't transfer is that the initialization of the neural network is an important part of the training process, which will have an important impact on the performance, convergence, and convergence speed of the model. Random initialization and stochastic gradient descent can cause the objective function to find a new local minimum (sometimes adjacent local minimum, resulting in slightly higher transferring attack performance, e.g., 68\%), which means that the resultant model is different each time. This result shows that the encoder, as well as the host model, can't be exactly reproducible with different random seeds for initialization, not to mention that it is almost impossible for an attacker to get the same architecture, dataset, and hyper-parameters. For more results of the transferring attack, see Figures in Appendix \ref{appendix_trans}.

%We also implement this attack on VGG-13 and ResNet-34, see Figures in Appendix B.

The above results hint at an advantage of learned encoders: unlike static watermark generation algorithms, they can employ a unique watermark generation strategy each time. Our IPP framework undoubtedly performs remarkable unforgeability against the fraudulent claims of ownership.

\begin{table}[t]
\centering
\caption{Robustness for  model fine-tuning: the trend of watermarking accuracy with fine-tuning epochs}
\setlength{\tabcolsep}{2mm}{
\begin{tabular}{|c|c|c|c|c|c|}  
\toprule
epochs  &V-13&V-16&R-18&R-34 & PreActR-18 \\ 
\midrule
0& 95.75\% &96.50\%& 97.00\%& 93.40\% & 91.75\%\\ 

10& 92.50\%  &95.50\%& 92.00\% & 90.70\%  & 90.75\%\\ 

20& 90.25\% &95.25\%& 91.75\%& 89.75\%  &90.00\%\\ 

30& 90.00\% &95.50\%& 90.50\%& 88.75\% &89.25\%\\ 

40& 90.00\%  &95.20\%& 89.75\%& 88.50\% &88.50\%\\ 

50& 90.00\%  &95.75\%& 89.50\%& 87.75\% &88.25\%\\

60&  90.00\%  & 95.25\% &89.00\%& 87.00\% &88.25\%\\

70&  90.15\%  & 95.00\% & 88.00\% & 86.75\% &88.00\%\\

80&  90.00\%  & 95.25\% & 87.75\%& 86.25\% &87.50\%\\

90&  90.25\%  & 95.50\%   & 87.50\% & 85.50\% &87.50\%\\

100& 90.00\%  &  95.50\%  & 87.50\% & 84.75\% &87.00\%\\
\bottomrule
\end{tabular}}
\label{robustness}
\end{table}

\subsection{Feasibility}
We consider two aspects of the feasibility: robustness and functionality. The purpose of robustness is to measure whether our framework is robust to the model modification, and the purpose of functionality is to measure whether our framework can associate the host model with the author's identity.

\textbf{Robustness}\label{robust}
Training a high-performance model from scratch requires a lot of resources, and insufficient datasets can significantly affect the performance of the model. Fine-tuning is a common strategy in practice, and we can fine-tune pre-trained models to complete new tasks or achieve higher performance. Therefore, an attacker can fine-tune the stolen model with fewer new datasets to obtain a new model that inherits the performance of the stolen model but is also different from the stolen model.

In this experiment, we divide the test set into two halves. The first half (80\%) is used for fine-tuning pre-trained models while the second half (20\%) is used for evaluating new models. We use watermark (key samples) accuracy of new models to measure the robustness of our framework for modifications caused by fine-tuning. Table \ref{robustness} shows that even after 100 epochs, our framework still has a high accuracy of all the models (only drop 9.50\% in the worst case). The reason behind this is fine-tuning can lead to an adjacent or same local minimum, which means weights of the model don't change significantly. This result indicates our framework can perform remarkable robustness against model modification. Note that, we don't consider such modification that can cause significant side effects on primary tasks, which leads to a totally different model. We agree with \cite{GP18} that the "proof-of-authorship" during actual model usage and "proof-of-origin" of a model should be two different problems.

\textbf{Functionality}
The functionality requires that our IPP framework shall clearly associate the host model with the author's identity. In a highly competitive global marketplace, the exclusive logo is the most widely used, the most frequent and the most important element in the process of corporate image transmission. However, none of the existing methods take it into account, and they do not closely associate the host model to be protected with the identity of the individual or organization. Essentially, we take advantage of the intrinsic over-fitting of the neural networks to achieve the goal, where we turn the weakness into a strength. Over-fitting is a modeling error that occurs when a function is too closely fit for a limited set of data points. In our framework, the host model is over-fitting for the key samples generated by encoder, and the encoder is over-fitting for the exclusive logo. This indicates that only the logo that participates in the training will lead a watermarked DNN to exhibit the desired behavior.
That is to say, our IPP framework can clearly associate the host model with the author's identity.

To the best of our knowledge, we are the first to take the usage of the logo into consideration and successfully design and implement our IPP framework. Furthermore, our framework also meets the requirements of white-box and black-box.

\begin{table}[t]
\centering
\caption{A summary of all methods meets the requirements}
\setlength{\tabcolsep}{2mm}{
\begin{tabular}{l|c|c|c|c|c|c}  
\toprule
Requirements   &\cite{ABCPK18,RCK18.0}&\cite{GP18}&\cite{NSn}&\cite{UNSS17}&\cite{ZGJWSHM18}&Ours   \\ 
\midrule
Fidelity& $\surd$&$\surd$ &$\surd$&$\surd$&$\surd$ &$\surd$  \\ 
\midrule
Effectiveness&$\surd$  &$\surd$ &$\surd$&$\surd$&$\surd$ &$\surd$    \\ 
\midrule
Integrity&$\surd$  &$\surd$ &$\surd$&$\surd$&$\surd$ &$\surd$  \\ 
\midrule
Security& &&$\surd$&$\surd$&&$\surd$  \\ 
\midrule
Legality& &$\surd$ &&$\surd$&&$\surd$  \\ 
\midrule
Feasibility&&$\surd$ &&&$\surd$ &$\surd$\\
\bottomrule
\end{tabular}}
\label{conclusion}
\end{table}

\section{Conclusion and Future Work}
In this paper, we propose the first blind-watermark based IPP framework aiming to generate the key samples of which the distribution is similar to the original samples. We successfully design and implement our IPP framework on two benchmark datasets and 15 popular deep learning models. We conduct extensive experiments to show that our framework is adequate to verify the ownership without significant side effects on primary tasks and achieves a remarkable performance on undetectability against evasion attack and unforgeability against fraudulent claims of ownership. Besides, our framework shows remarkable robustness against model modification. Lastly, we are the first to take the usage of the logo into consideration and establish a clear association between the model and the creator's identity. In briefly, our IPP framework meets all the requirements summarized in Table \ref{conclusion}.

For future work, we plan to protect the intellectual property of the speech recognition model. We expect this work to be expanded to other forms of deep learning, e.g., recurrent neural networks, etc.

\section{Acknowledgments}
We would like to thank the anonymous reviewers for their comments on previous drafts of this paper. This work is partially supported by National Natural Science Foundation of China (91546203), the Key Science Technology Project of Shandong Province (2015G GX101046), Major Scientific and Technological Innovation Projects of Shandong Province, China (2018CXGC0708, 2017CXGC0704).
\bibliographystyle{ACM-Reference-Format}
%\bibliography{sample-base}
\bibliography{main}
%\begin{appendices}
\appendix
\section{The architectures of encoder and discriminator}\label{appendix_arch}
We first denote the abbreviation of Conv2d, ConvTranspose2d, Batch Normalization, and LeakyReLU as Conv, ConvT, BN, and LReLU. We also define a UnetSkipConnectionBlock, which is adapted from the architecture of U-net \cite{UNET15}, with $4\times4$ kernel, stride 2 and padding 1. For reproducibility purposes, our code is available at \url{https://github.com/zhenglisec/Blind-Watermark-for-DNN}.
\subsection{Encoder}
Given input ordinary sample of $3\times H \times W$ and logo of $3\times H\times W$, the  encoder applies 2  ConvT-BN-LReLU  blocks  with 3$\times$3  kernel,  stride  1,  padding  1,  and $3$ output  filters  each for prepossessing the ordinary sample and logo respectively, resulting in $3 \times H \times W$ image activation feature. The two feature maps are concatenated to a tensor of size $6 \times H \times W$ where the logo feature map is dispersed throughout all bits in the ordinary feature. 
Then the encoder applies 5 UnetSkipConnectionBlocks with 64, 128, 256, 512, 512 filters each (from outside to inside) to reduce the number of channels from $6$ to $3$. Finally, we get the output of the encoder, the key image of size $3 \times H \times W$.

As for the architecture of encoder for mnist, due to the input of size $1 \times H \times W$, the encoder first applies 2 ConvT-BN-LReLU blocks for prepossessing the ordinary sample and logo respectively, resulting in $3 \times H \times W$ image activation feature. The two features are concatenated into a 6-channel tensor as the input of the next block with 3$\times$3  kernel, stride 1, padding 1. The block is composed of 3 Conv-BN-LReLU with 64 filters, and Conv-Tanh with 1 output filters. Then the 6-channel tensor is reduced to 1-channel tensor. Finally, we get the output of the encoder, the key sample of size $1 \times H \times W$.

\subsection{Discriminator}
The discriminator accepts the ordinary sample and the key samples as input, respectively. We first define the LinearBlock which is composed of a fully connected layer and LReLU. The discriminator applies 2 LinearBlocks with 64 and 32 output filters each, and particularly a fully connected layer followed by the Sigmoid activation to outputs a binary classification $p$ for a given image, i.e., either an ordinary sample or key sample. 

\section{The performance of transferring attack on VGG-13 and ResNet-34}\label{appendix_trans}
%\subsection{The performance of transferring attack on VGG-13 and ResNet-34}
%As depicted in Figure \ref{heatmapvgg} and Figure \ref{heatmapresnet}, the high accuracy of the attack results is listed at the diagonal of the Figure, which shows that the key samples can only induce their corresponding model to pre-defined labels. These results show that the learned model can’t be perfectly reproducible with different random seeds for initialization.
\begin{figure}[htbp]
\centerline{\includegraphics[width=7cm, height=6cm]{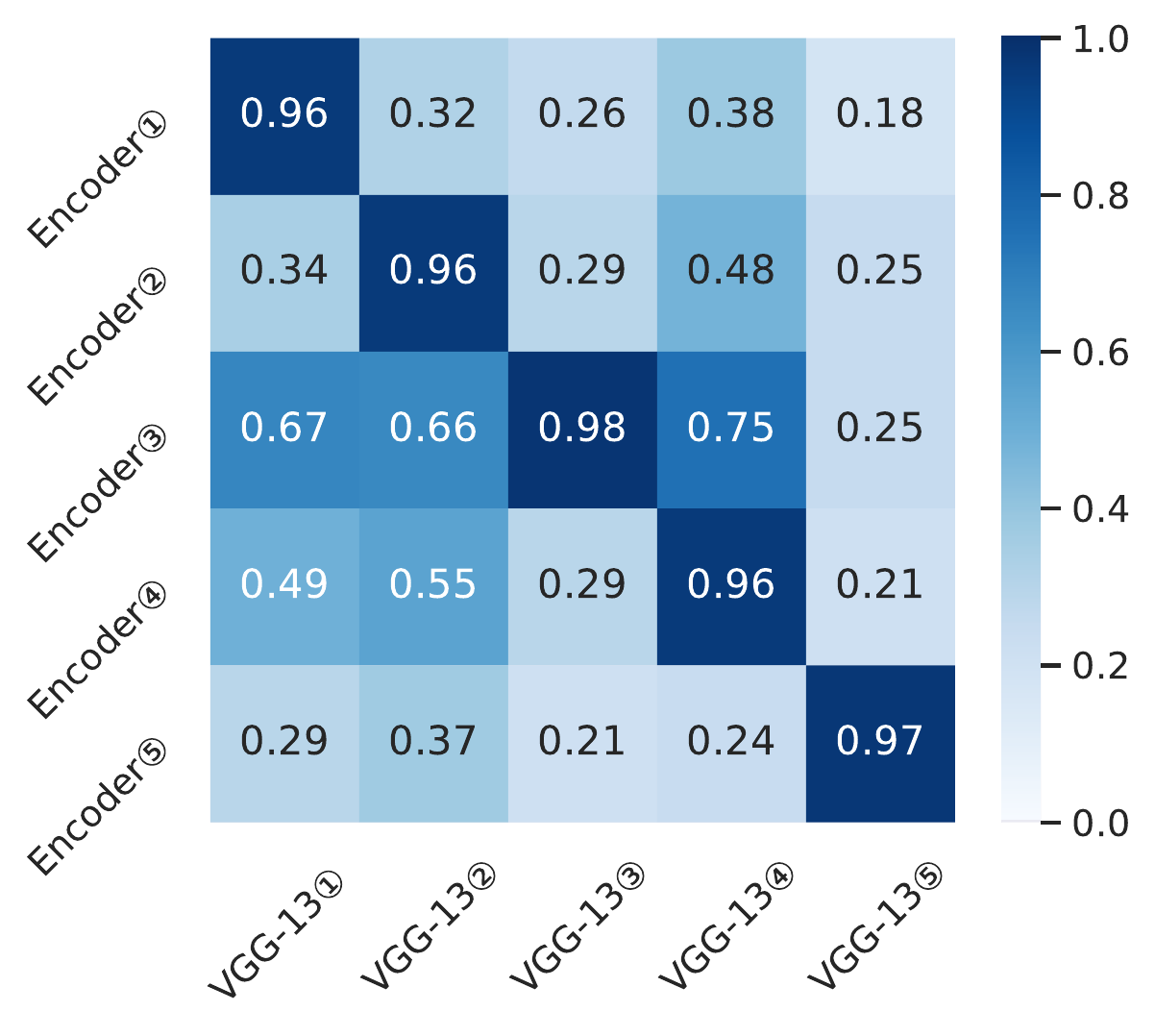}}
%\caption{The performance of transferring attack on VGG-13}
\centerline{\includegraphics[width=7cm, height=6cm]{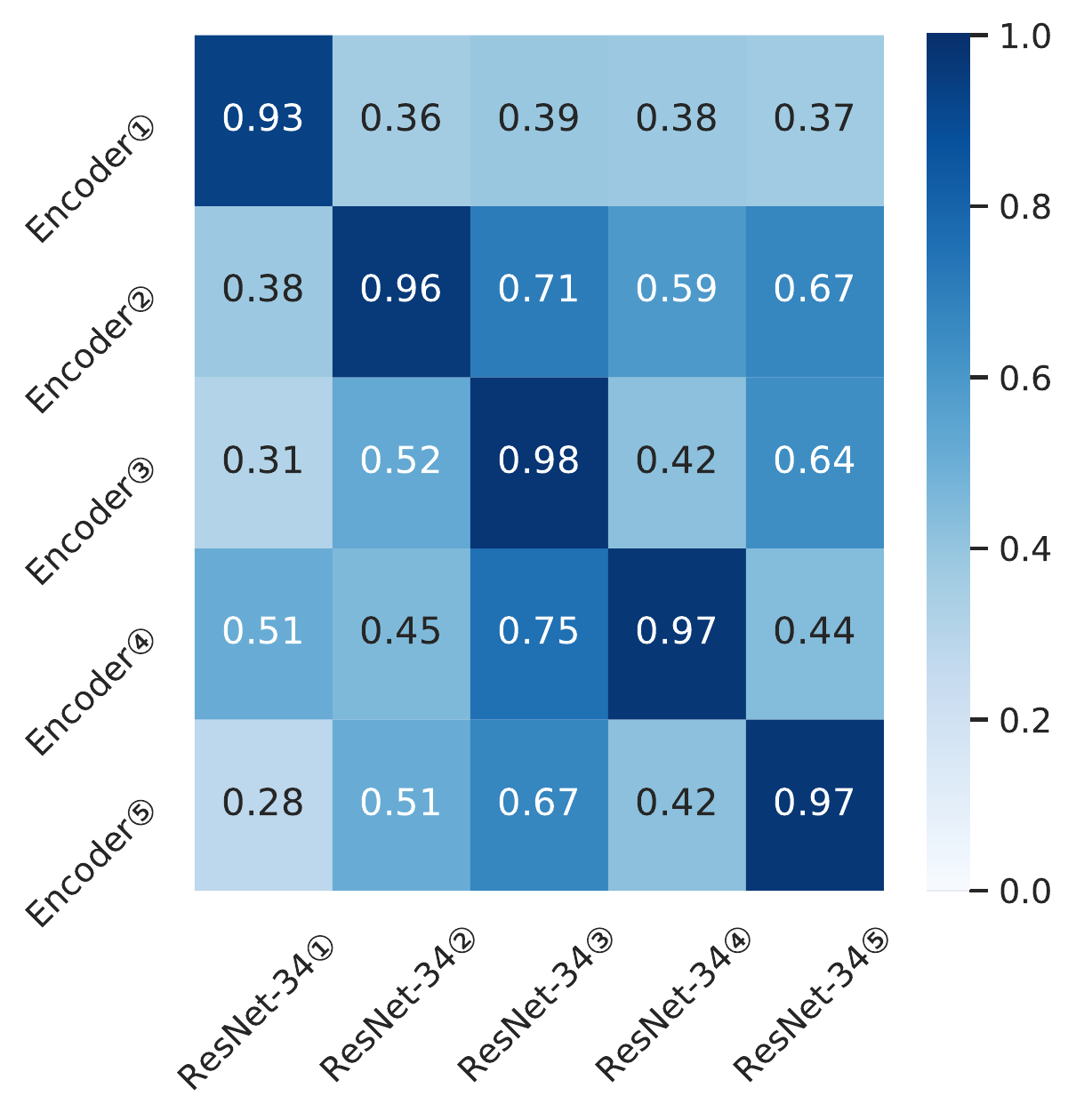}}
%\caption{The performance of transferring attack on ResNet-34}
\end{figure}
%\end{appendices}
\end{document}